\documentclass[twocolumn,english]{revtex4-1}

\usepackage[T1]{fontenc}
\usepackage{xcolor}
\usepackage{float}
\usepackage{calc}
\usepackage{amsmath}
\usepackage{amsthm}
\usepackage{graphicx}
\graphicspath{{figures/}}
\usepackage{algorithm}
\usepackage{algpseudocode}
\usepackage{overpic}
\usepackage{multirow}
\usepackage{tikz}
\usepackage{pgfplots}
\pgfplotsset{compat=newest}
\usepgfplotslibrary{patchplots}
\usepackage{subfig}
\makeatother

\usepackage{babel}
\begin{document}


\title{Unsupervised learning of control signals and their encodings in \emph{C. elegans} whole-brain recordings}

\author{Charles Fieseler$^*$, Manuel Zimmer$^\dag$$^{**}$ and J. Nathan Kutz$^{***}$}

\affiliation{$^*$Department of Physics, University of Washington, Seattle, WA 98195\\
$^\dag$Department of Neurobiology, University of Vienna, Althanstrasse 14, 1090 Vienna, Austria\\
$^{**}$Research Institute of Molecular Pathology (IMP), Vienna Biocenter (VBC), Campus-Vienna-Biocenter 1, 1030 Vienna, Austria\\
$^{***}$Department of Applied Mathematics, University of Washington, Seattle, WA 98195}

\keywords{C elegans $|$ Control $|$ DMD $|$ ...} 

\begin{abstract}
	Recent whole brain imaging experiments on \textit{C. elegans} has revealed that the neural population dynamics encode motor commands and stereotyped transitions between behaviors on low dimensional manifolds. 
		Efforts to characterize the dynamics on this manifold have used piecewise linear models to describe the entire state space, but it is unknown how a single, global dynamical model can generate the observed dynamics.
	Here, we propose a control framework to achieve such a global model of the dynamics, whereby underlying linear dynamics is actuated by sparse control signals.
	This method learns the control signals in an unsupervised way from data, then uses {\em Dynamic Mode Decomposition with control} (DMDc) to create the first global, linear dynamical system that can reconstruct whole-brain imaging data.
	These control signals are shown to be implicated in transitions between behaviors.
	In addition, we analyze the time-delay encoding of these control signals, showing that these transitions can be predicted from neurons previously implicated in behavioral transitions, but also additional neurons previously unidentified.
	Moreover, our decomposition method allows one to understand the observed nonlinear global dynamics instead as linear dynamics with control.
	The proposed mathematical framework is generic and can be generalized to other neurosensory systems, potentially revealing transitions and their encodings in a completely unsupervised way.
\end{abstract}



\maketitle

\section{Introduction }
The nematode {\em Caenorhabditis elegans} (\emph{C. elegans}) is an ideal model organism {for probing  the relationship between structure and function in neuronal networks} as it is comprised of only 302 sensory, motor, and inter-neurons whose stereotyped synaptic connections (i.e. its connectome) are known from serial section electron microscopy~\cite{White_paper, cook2019whole, jarrell2012connectome}.  
Indeed, \emph{C. elegans} is perhaps the simplest organism to display many of the hallmark features of high-dimensional networked biological systems, including the manifestation of low-dimensional patterns of activity associated with functional behavioral responses \cite{kato2015global}.  
Thus the nervous system must reduce the high-dimensional representation of environmental stimuli into much lower dimensional representations of motor commands~\cite{roberts2016stochastic,liu2017functional,kutz2016dynamic,kunert2017multistability,Fieseler_paper1, kato2015global, kawano2011imbalancing}.  
Low dimensional representations have been separately considered in posture (behavioral) analysis \cite{stephens2008dimensionality,stephens2011emergence} as well as in previous analysis of calcium imaging data \cite{kato2015global,nichols2017global}.
{
These representations can be used to characterize the evolution of both postures \cite{stephens2011emergence} and neuron population dynamics \cite{brennan2019quantitative, linderman2019hierarchical}.
}
%
%
%
In this work, we exploit emerging whole-brain imaging recordings to posit a data-driven model of neurosensory integration in \emph{C. elegans}, showing that a global linear framework, with the addition of internally generated control signals, explains and reproduces much of the activity of the network.

It has long been observed that \emph{C. elegans} produces a small number of stable discrete behaviors (e.g. forward and backward motion, and turns), and that these behaviors change both spontaneously \cite{roberts2016stochastic} and very quickly in response to external stimuli \cite{zimmer2009neurons, chalasani2007dissecting, chiba1990developmental} or stimulation of even a single neuron \cite{kocabas2012controlling,leifer2011optogenetic,nagel2005light}.
A potential dynamical systems explanation for this observation is that of discrete behaviors as fixed points on an underlying manifold with some transition signals that move the system between them. 
{
A purely linear dynamical system of the form ${\bf x}_{k+1} = {\bf A}{\bf x}_k$, cannot produce the observed multiple fixed points, where ${\bf x}_k$ is the data at time point $k$ and the matrix ${\bf A}$ maps the state one step into the future .}
However, piecewise methods, like switching (hybrid) linear dynamical systems \cite{linderman2014discovering, linderman2016recurrent, costa2019adaptive, linderman2019hierarchical}, circumvent this by segmenting the dynamics into patches with different dynamics (and thus different fixed points) in each patch. 
An alternate method uses different phase loops and the phase along them to predict behavior, producing conserved dynamics in a special phase space \cite{brennan2019quantitative}.
Recent efforts have also attempted to explicitly model neuronal and synaptic dynamics to approximate biophysical models of the nervous system~\cite{kunert2014low, mujika2017modeling, costalago2018emulation, izquierdo2018role, gleeson2018c302, Fieseler_paper1, boyle2012gait}, but this has currently been limited to subsets of neurons and has moreover had difficulty capturing multiple behaviors.
This work instead focuses on how a single, global, {dynamical system model} with simple and interpretable additions can capture the nonlinear dynamics via appropriate framing as a control problem.

The recent availability of real-time calcium imaging data allows for a neuron-level, data-driven approach.
{The goals of a} full model of \emph{C. elegans} neural activity are to describe how multiple states are produced in a single network, and how dynamics operating at multiple scales are integrated to produce the states and transitions between them.
{Our work mathematically frames this biological problem in the context of control theory by learning both the dynamics and its encoding.}
{The dynamics portion is implemented using the recently developed} data-driven method of {\em Dynamic Mode Decomposition with control} (DMDc) \cite{kutz2016dynamic,proctor2016dynamic}, demonstrating that additional nonlinearities are not needed to describe many of the interactions in the system.
{We additionally extend this method to handle unsupervised learning of control signals as described in the Methods Section, allowing generalizability to new complex systems and testable hypotheses in this system as discussed in the Results Section.
The encoding portion is implemented using sparse variable selection methods and the novel elimination pathway of the encoding (see Methods), revealing the timescales and locations where these transition signals are encoded (see Results and Supplementary Material).
}
We provide code written in MATLAB \cite{this_code_github} for a full analysis pipeline that uses raw data and, if available, external behavioral labels to discover both the intrinsic dynamics, the effects of control on the state of the system, and {the encoding of the control signals.}

\begin{figure*}[t]
\includegraphics{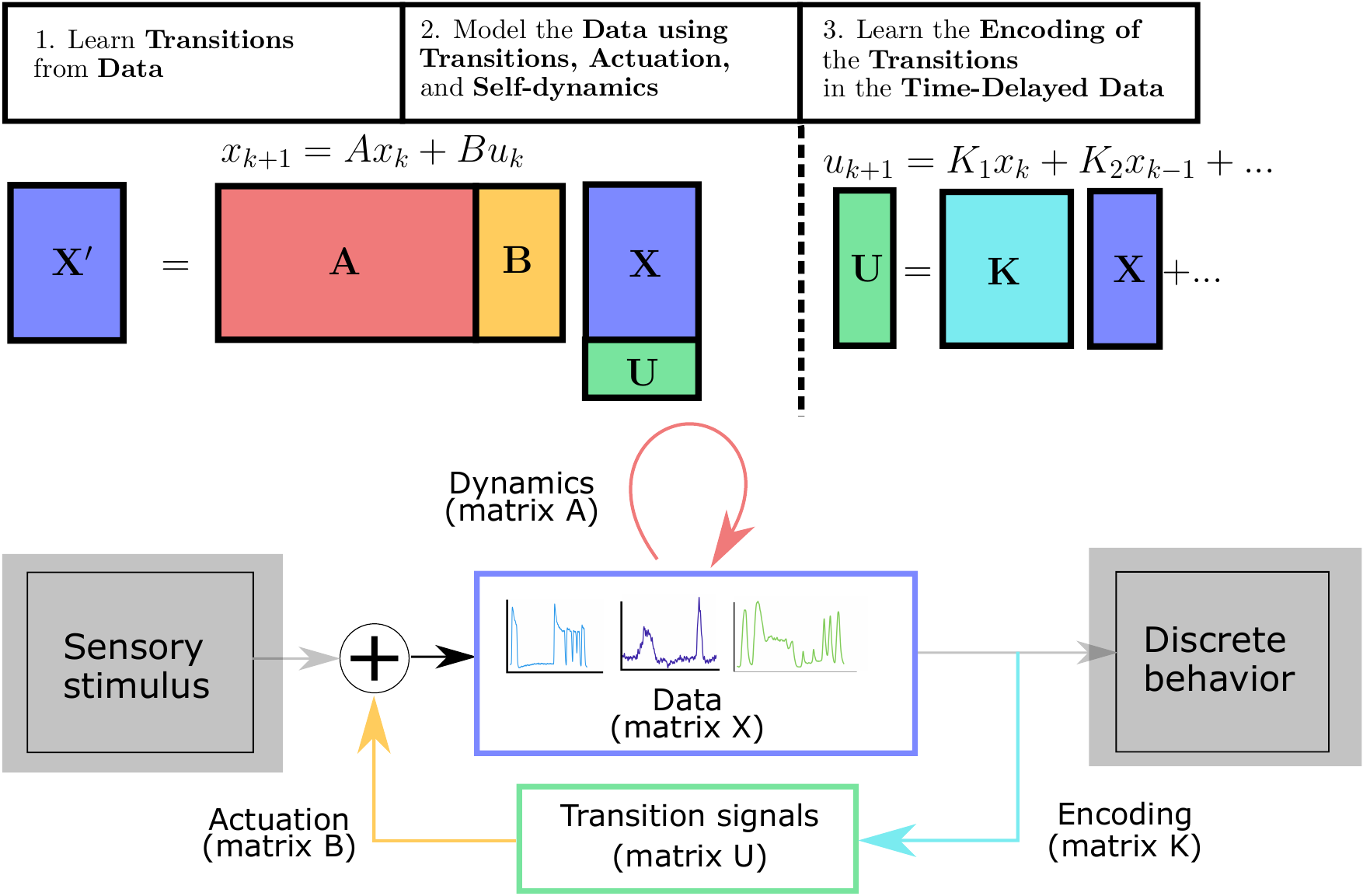}
\caption{A 3-step framework for modeling neurosensory integration. 1) Transition signals are learned from data with an assumption of linear dynamics. 2) A DMDc model is learned which uses dynamics, transition signals, and actuation. These are global models, and are capable of reconstructing much of the data dynamically from an initial state. 3) Where and at what timescales control signals are encoded in the neural activity is studied using sparse linear models.
{Bottom: the sensory-computation-behavior pathway. Each term in the above equations can be freely translated into this biological process. Transitions (Green) actuate neurons via their own connectivity (Yellow). Neuron traces (Blue) evolve according to intrinsic dynamics (Red), and also encode (Light Blue) the transition signals (Green).}
\label{fig:generalization_path}}

\end{figure*}

\begin{figure}
\includegraphics{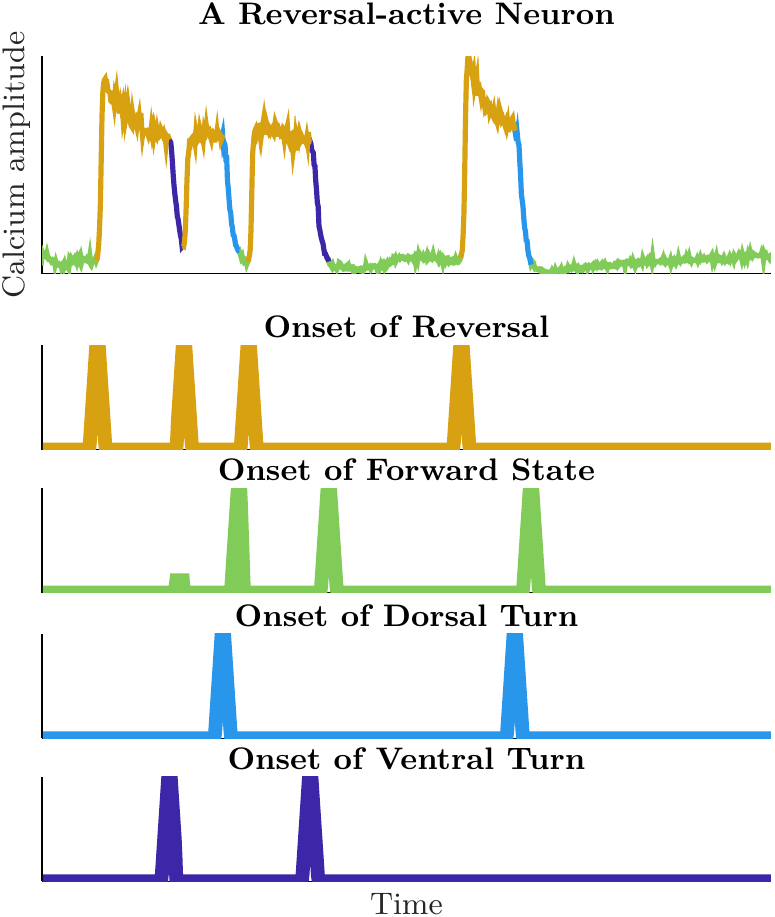}
\caption{Transition signals in \emph{C. elegans}:
Top: A calcium imaging trace of a neuron connected with the discrete reversal behavior. Behavioral labels are determined by experimentalists, as described in \cite{kato2015global}. Green=Forward; Yellow=Reversal; Dark Blue=Ventral Turn; Light Blue=Dorsal Turn.
Below: These labels can be reframed as ``onset'' signals, and are characteristically sparse in time.
\label{fig:Control_framing_intro}}
\end{figure}

\section*{Data-Driven Methods}

Our analysis relies on two established mathematical methods: DMDc and sparse optimization.  A  brief summary of each is given below.

\subsection*{Dynamic Mode Decomposition with control}

Our data-driven strategy is based upon the {\em dynamic mode decomposition} (DMD).  DMD provides a linear model for the dynamics of the state space ${\bf x}_j={\bf x}(t_j)$ given the data matrices constructed using temporal snapshots of the system, ${\bf X} = \left[  {\bf x}_1 \,\,{\bf x}_{2}  \,\, \dots \,\, {\bf x}_{m-1}   \right]$ and ${\bf X}' = \left[  {\bf x}_2 \,\,{\bf x}_{3}  \,\, \dots \,\, {\bf x}_{m}   \right]$.  Specifically, it finds the best fit linear model 
\begin{equation}
  {\bf X}' = {\bf A}{\bf X} \label{eq:dyn3a}
\end{equation}
There are a number of variants for computing ${\bf A}$~\cite{kutz2016dynamic}, with the {\em exact DMD} simply positing ${\bf A}={\bf X}' {\bf X}^\dag$ where $\dag$ denotes the Moore-Penrose pseudo-inverse. 

DMDc \cite{proctor2016dynamic} capitalizes on all of the advantages of DMD and provides the additional innovation of being able to disambiguate between the underlying dynamics and actuation signal ${\bf u}_j={\bf u}(t_j)$.  For a matrix of input signals ${\bf U} = \left[  {\bf u}_1 \,\,{\bf u}_{2}  \,\, \dots \,\, {\bf u}_{m-1}   \right]$, DMDc regresses instead to the linear control system
\begin{equation} {\bf X}' = {\bf A} {\bf X} + {\bf B} {\bf U}. \label{eq.dyn3b} \end{equation}  
Note that DMDc uses only snapshots in time of the state space and control input, making it compelling for systems whose governing equations are unknown. 
The DMDc equation is graphically represented in Fig.~\ref{fig:generalization_path}.
The governing matrices (\textbf{A} and \textbf{B}) along with the control signal (\textbf{U}) produce a predictive model, such that the state of the system far in the future can be predicted.
For instance, the third time step can be estimated from the first via:
\begin{equation} 
{\bf x}_3 = {\bf A} ( {\bf A} {\bf x}_1 + {\bf B} {\bf u}_1 ) + {\bf B} {\bf u}_2 \label{eq:predict_1} 
\end{equation} 
%

\subsection*{Learning control signals via sparse optimization}

\begin{figure*}
\includegraphics{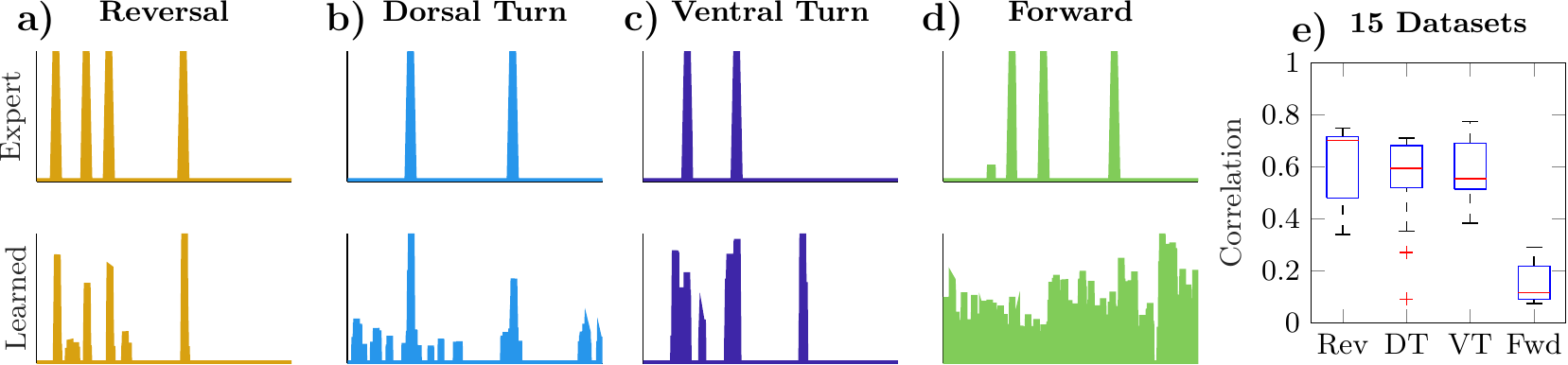}

\caption{Control signals can be learned from data via algorithm \ref{alg:sparse_residual_analysis}. 
a-d) The onset of well-known states as determined by experts (above) and as learned (below). 
{All signals are normalized to have a maximum of 1.0.}
e) Correlation between expert and learned signals across 15 individual datasets. Reversals (Rev), Dorsal (DT) and Ventral Turns (VT) are consistently learned, but Forward state (Fwd) onsets are never significant.
\label{fig:Unsupervised_controllers}}
\end{figure*}

The DMDc algorithm requires knowledge of the linear control signals ${\bf U}$.
Expert-identified state labels and an example neuron that displays strong state-dependent behavior are shown in figure \ref{fig:Control_framing_intro}.
However, these are only available because of the decades of \emph{C. elegans} experimental work identifying 1) discrete behavioral states and 2) the command neurons for each activity.
For new organisms, and in order to generate hypotheses about potential new states in \emph{C. elegans}, the unsupervised problem, i.e. learning the signal directly from data, is of critical interest.

DMDc (\ref{eq.dyn3b}) can be thought of as an error minimization problem
over the dynamics and actuation matrices, ${\bf A}$ and ${\bf B}$. 
If the control signal is unknown, the minimization must be extended to the control signal ${\bf U}$ itself. 
However, there is now a trivial solution where the control signal dominates the model: ${{\bf X}_2={\bf BU}}$ with ${{\bf A}=0}$.
For this reason, an assumption must be made about the control signals.
In this case, the statement that these signals are sparse is directly biologically interpretable, and means that the transitions between states should be rare as a percentage of frames.
This ``sparsity constraint'' can be expressed in a mathematically precise way using the $\ell_0$ norm:
\begin{equation} 
\min_{{\bf A},{\bf B},{\bf U}}\left[ \left|\left|{\bf A}{\bf X}_{1}+{\bf B}{\bf U}-{\bf X}_{2}\right|\right|_{2} + \lambda \left|\left|{\bf U}\right|\right|_{0} \right]
\label{eq:dmdc_optimization_sparse}
\end{equation}
Directly solving this optimization problem is extremely difficult, although there are efficient algorithms in certain cases \cite{jewell2018exact}.  More recently, a convex relaxation of the $\ell_0$ to an $\ell_1$ norm is often solved \cite{donoho2006most}, though this has been recently shown to lead to errors in its selection pathway \cite{su2017false}.
We use a different approximation, the sequential least squares thresholding algorithm~\cite{brunton2016sparse}, which has been shown to converge to the minima of the original $\ell_0$ problem \cite{zhang2018convergence,zheng2018unified}.
The code is outlined in algorithm \ref{alg:sparse_residual_analysis} and more detail is given in the supplement.
The matrix ${\bf U}$ in this algorithm is additionally constrained to be positive, for better interpretability as ``on'' transition signals.

\begin{algorithm}[H]
\begin{algorithmic}[1]
\Procedure {LearnControllers}{$r$}
\State ${\bf U}_0:= InitializeU(\text{r})$
\State ${\bf S}:= InitializeSparsityPattern({\bf U}_0)$
\For{$i\gets 1, MaxIter$}
	\State ${\bf A}, {\bf B} = SolveAB({\bf X}, {\bf U}_{i-1})$ 
	\State ${\bf U}_i = SolveU({\bf X}, {\bf A}, {\bf B})$
	\State ${\bf S} = UpdateSparsityPattern({\bf S}, {\bf U}_i)$
	\State ${\bf U}_i({\bf S}) = 0$
\EndFor
\EndProcedure
\end{algorithmic}
\caption{Unsupervised Learning of Control Signals \label{alg:sparse_residual_analysis}}
\end{algorithm}

\subsection*{Variable selection via sparse linear models}

If internally generated control signals are present, then there are two {possibilities}: they are random and fundamentally unpredictable, or they are encoded in the network. 
{We explore the degree to which these signals are encoded using sparse variable selection algorithms and \emph{time-delay embedding,} where data from further in the past is utilized.
Mathemaically, this is written as}:

\begin{equation}
{\bf U} = {\bf K}_1 {\bf X}_1 + {\bf K}_2 {\bf X}_2 + ... \label{eq:encoding}
\end{equation}

There are multiple methods that are often used to perform this variable selection task \cite{tibshirani1996regression}.
However, these methods may make mistakes in their selections \cite{su2017false}, and in general it is unclear how unique the selection is. 
The behaviors of \emph{C. elegans} have been well studied, and each onset is associated with well-known neurons. 
Variable selection methods will almost certainly discover these well-known neurons, but by exploring further in the ``elimination path'',  less obvious encodings can be discovered.
Algorithmically, this is the sequential removal {or ablation} of the most important neuron for all time delays, and then a re-fitting of the sparse model.
If the quality of the reconstruction does not degrade along the elimination path, the signal (${\bf U}$) must be distributed throughout the data (${\bf X}$).
The quality of signal reconstruction is defined here as the number of false positives and false negatives in the reconstructed signal.
Event detection is defined as a {minimum} number of frames above a hard threshold, as shown in Fig. \ref{fig:variable_selection} and discussed in the supplement.


\begin{figure*}[t!]
\includegraphics{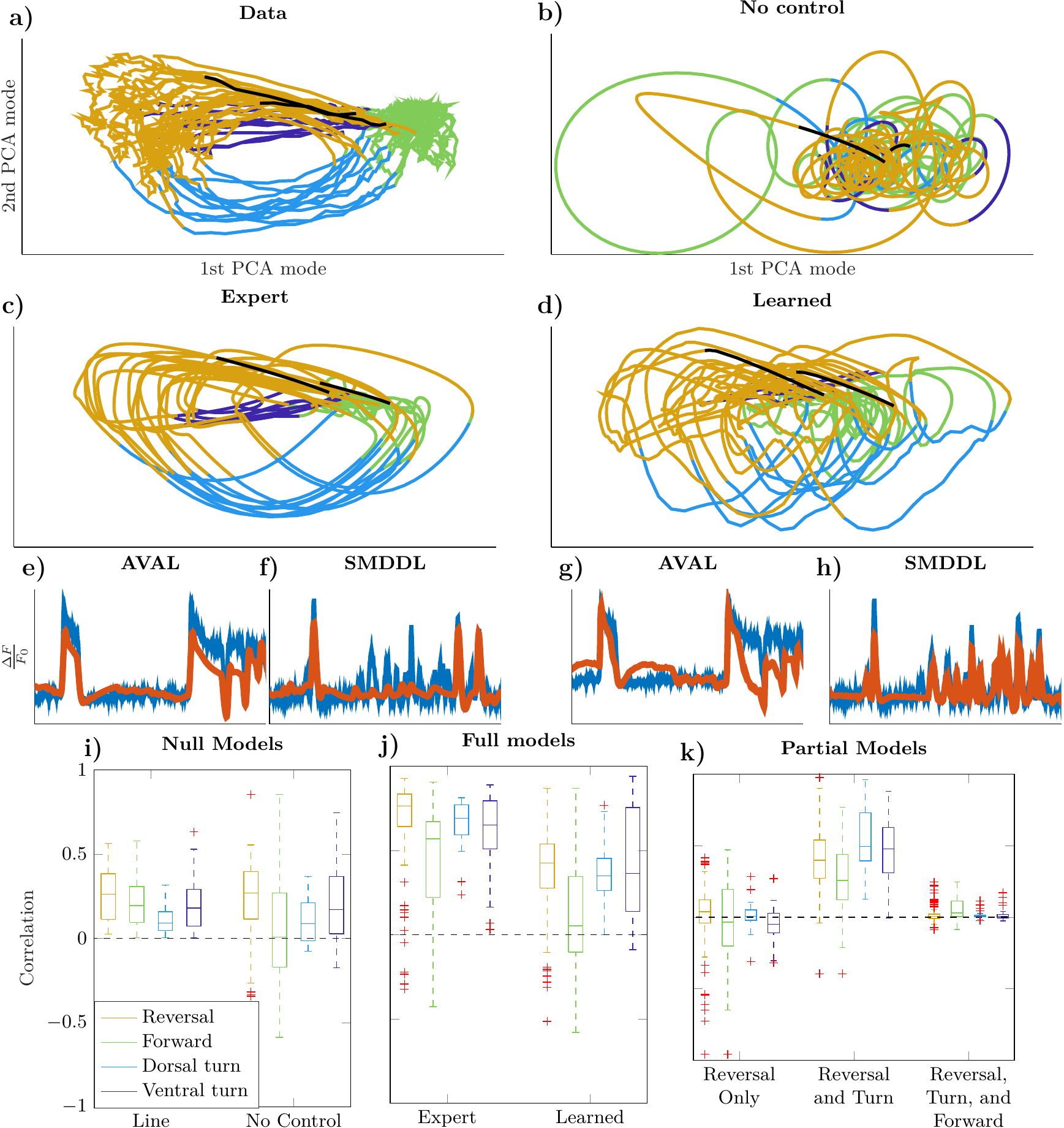}
\caption{
2d PCA projections of a) data, b) an uncontrolled ``null'' model, c) a ``supervised'' model using expert-determined control signals, and d) an ``unsupervised'' model that uses control signals learned via algorithm \ref{alg:sparse_residual_analysis}. The governing equations matrices in are all learned from data, either uncontrolled (b, Equation \ref{eq:dyn3a}) or controlled (c-d, Equation \ref{eq.dyn3b}).
These data are color-coded by state: Black for Unknown, Yellow for Reversal neurons, Green for Forward, and Light (Dark) Blue for Dorsal (Ventral) turns.
e-f) Example neuron datasets with reconstructions from the supervised model. A reversal-active (AVAL) and a Dorsal-Turn-active neuron (SMDDL) are shown.
g-h) The same neurons shown with reconstructions from the unsupervised model.
i-k) Correlations across datasets between data and reconstructions, split up into 4 different neuron groupings for interpretability. 
i) Baseline null models. The left-hand side is a straight line fit to a neural trace, {i.e. how well pure drift can explain the signal}. The right-hand side corresponds to the uncontrolled model in panel (b), and is generally worse than a straight-line fit.
j) Full models with either expert/supervised control signals, (c) above, or learned/unsupervised control signals, (d) above. For each neuron grouping the expert signals produce significantly better fits.
k) Partial supervised models, as more signals are added. For the Reversal (left-hand side) set, a ``baseline'' of a straight-line fit is subtracted. Shown are additive improvements, i.e. how much better each partial model is than the one immediately to the left.
\label{fig:Reconstructions}
}
\end{figure*}

\begin{figure*}[t!]
\includegraphics{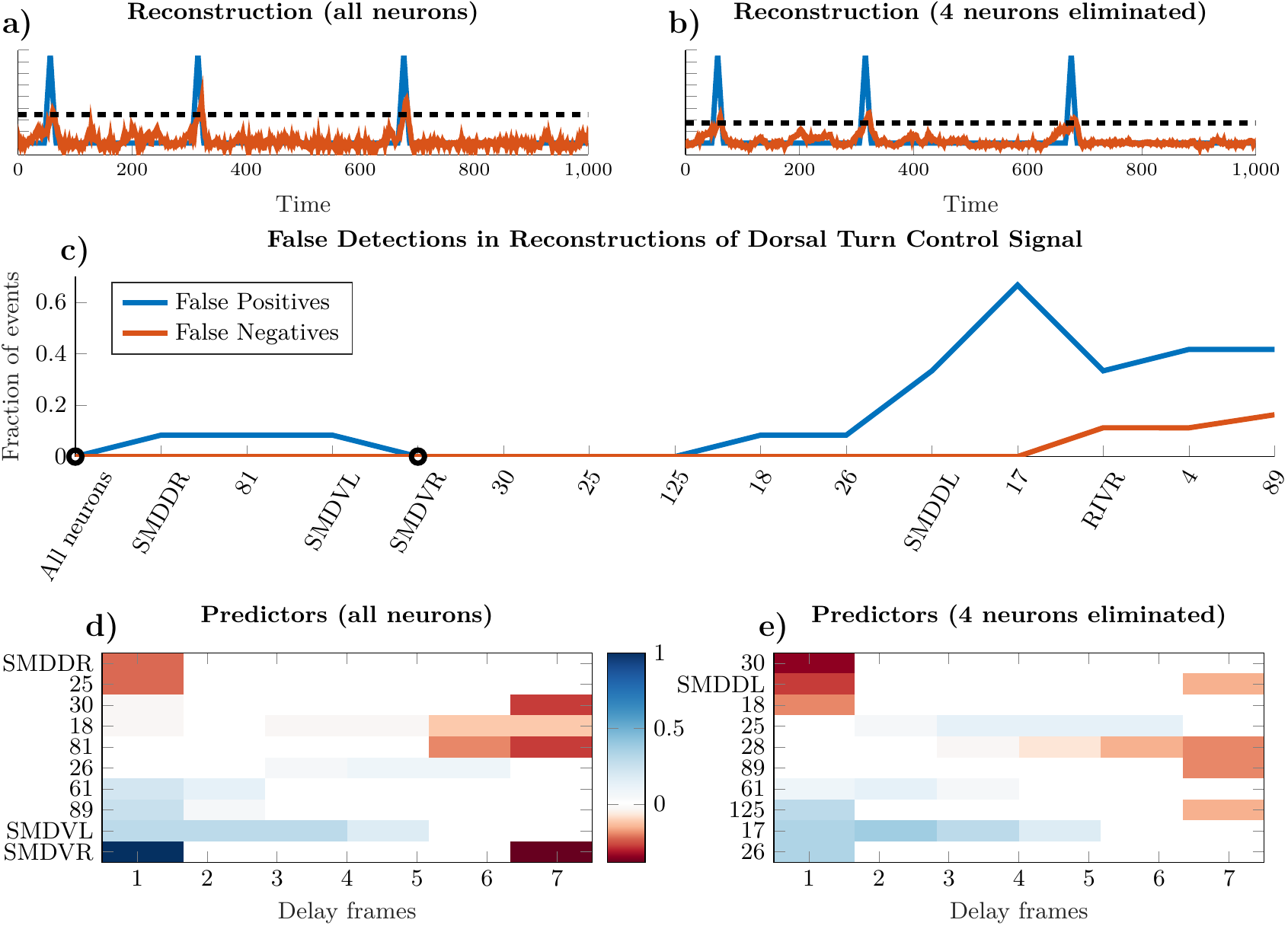}
\caption{
a-b) Control signal reconstructions via linear encoding on the data including time delays, with all neurons (a) or 4 neurons removed (b). {The removed neurons are: SMDDR, 81, SMDVL, and SMDVR as shown on the x axis of (c). Event detection is determined via a hard threshold for each signal (dotted line). See supplement for more discussion of this threshold.
c) Neurons are eliminated in order of the largest magnitude given to them by the linear model. The number of false detections increases significantly only after ~8 neurons have been removed.}
d-e) The weights given to the top 10 most important neurons for different iterations.
\label{fig:variable_selection}
}
\end{figure*}

\section*{Results}

\subsection*{Known transitions are discovered and characterized}

Experimentalists have long separated behavior into discrete categories through careful study of individual neurons. 
However, open questions remain about the number of behaviors that exist and how discrete they are.
Some works have posited up to six forward motion states and three reversal states, multiple turning subtypes, or even a continuum of behaviors \cite{szigeti2015searching, liu2018temporal}.
As Fig. \ref{fig:Unsupervised_controllers} shows, using unsupervised optimization three behavioral onsets can be discovered: Reversal, and Dorsal and Ventral turns.
In particular, the single Reversal onset signal for each individual suggests that this transition is fundamentally the same within individuals, with variability produced by activation amplitude but not a different direction in neuron-space.

However, in no individuals could a signal correlated to the onset of Forward motion be discovered.
{If this were a trivial state that displays no activity, a simple decay to a fixed point following a Turn state would be sufficient to achieve a good reconstruction the trajectory, even without an onset signal.
However, fast-scale behaviors are known experimentally to occur within this state \cite{kaplan2019nested}, and this complexity is reflected in the poorer reconstruction quality as shown in Fig. \ref{fig:Reconstructions}.j. 
Thus, although there is activity, its onset falls outside the sparsity assumptions of Eq. \ref{eq:dmdc_optimization_sparse}.}
Taken together, our results imply that for the onset of these behaviors, forward motion is more complex than reversals, meaning that it cannot be described as a simple ``on'' signal. 
{It is known that reversals and forward motion largely activate disjoint subnetworks of neurons \cite{zhen2015c}, but the effects of this separation remain unclear.}

\subsection*{A global, linear system with control reconstructs entire time series}

The manifold observed in \emph{C. elegans} neural dynamics cannot be described by a purely linear model due to the presence of multiple stable global behaviors, as shown in Fig. \ref{fig:Reconstructions}.b.  Specifically, linear models can only admit a single fixed state.
However, the majority of neurons can be reconstructed using our \emph{controlled, global, linear} dynamical system due to the sparse transition signals as shown in Fig. \ref{fig:Reconstructions}.c for expert hand-labeled signals and Fig. \ref{fig:Reconstructions}.d for signals learned from data.
Each time snapshot of this data is reconstructed analogously to equation \ref{eq:predict_1}, and then projected onto the two dominant PCA modes of the original data so that each panel in Fig. \ref{fig:Reconstructions}.a-d is in the same coordinate space.
Because this is a global linear model that uses a single framework for the entire state space, the need for additional nonlinear modeling can be constrained to particular groups of neurons and well-defined time windows.

In particular, across individuals the reversal class of neurons is captured very well by the supervised control signal as shown in Fig. \ref{fig:Reconstructions}.j and thus, up to encoding the transition signal itself, the relevant subnetwork does not appear to require nonlinearities.
This means that future efforts related to nonlinear modeling {may be most productive if they} concentrate on the small window of time during the onset of the behavior, instead of the entire neural trace where linear models {are sufficient}.
In addition, the type of nonlinearity required to more fully model this class of neurons is characterized: fast and short-lived spike-like activations.

Turns are also largely captured, as shown by the high correlation for the light and dark blue boxplots in Fig. \ref{fig:Reconstructions}.j.
The neurons involved in turning have a large number of smaller events, as shown in the SMDDL reconstruction Fig. \ref{fig:Reconstructions}.f; these do not lead to one of the four state transitions identified by experimentalists in this dataset, but may correspond to recently described fast time-scale states \cite{kaplan2019nested}.
However, the unsupervised method does pick up on these smaller events and reconstructs them well as in Fig. \ref{fig:Reconstructions}.h, but over all datasets there is much more variability as shown in Fig. \ref{fig:Reconstructions}.j.

The last group of neurons, those related to forward motion, has a very large variability of correlation between the data and reconstructions.
This implies that these neurons require nontrivial nonlinearities throughout the time series, {not just at the onset, for full reconstruction.
Although it is well known that different, dedicated subnetworks of neurons are active in forward and backwards motion \cite{zhen2015c}, the functional implications have not been fully modeled.
Some recent experimental work \cite{kawano2011imbalancing} characterizes an asymmetry between Forward and Reversal states as due to intrinsic bias towards the Forward state.
In addition and unlike the Reversal-active neurons that require only a simple ``on'' transition signal, the Forward-active neurons} may be continuously parametrized by speed, or contain additional behaviors like steering \cite{iino2009parallel}, tracking \cite{luo2006sensorimotor}, {or head casting \cite{kaplan2019nested}.
Moreover, Kaplan et al. show that many neurons exhibit diverse faster time-scale fluctuations particularly during forward states \cite{kaplan2019nested}.
Our work is consistent with these experimental results, and adds that this complexity is functionally different than that of the Reversal or Turn states.}

To further characterize the effects of the control signals on the ability of this framework to capture the neural dynamics, partial models were created with a subset of control signals.
{Partial models using cumulative subsets of the expert-labeled control signals} are shown in Fig. \ref{fig:Reconstructions}.k.
Adding Reversal-onset signals alone does not produce a model that captures the data better than a straight-line fit to the data, but the combination of Reversal and Turning signals is significantly better.
However, subsequent addition of Forward control signals is, remarkably, useless. 
{In summary, there are several related functional observations that further work may connect to physical differences in the Reversal and Forward neuronal subnetworks: the lack of discovery of sparse Forward onset signals, which is corroborated by the ineffectiveness of the experimentally known onset times, and the poor reconstruction of Forward-related neurons using linear dynamics.}

\subsection*{Transitions are encoded in previously {unidentified} neurons}

Having shown the control signals to contribute significantly to the reconstruction of the data, we reconstruct the control signals themselves using time-delay data matrices and sparse linear models as shown in step 3 of Fig. \ref{fig:generalization_path} according to equation \ref{eq:encoding}.
As described in \cite{nichols2017global, kato2015global}, each of the four interpretable transition signals shown in figure \ref{fig:Unsupervised_controllers} are hand-labeled using the activity of certain well-known neurons.
Thus, it is not surprising that these signals can be reconstructed from data when those well-known neurons are included. 
In particular, as they were used to define the Dorsal Turn behavioral states, like SMDVL/R which define Ventral Turns \cite{kaplan2019nested}, an excellent validation is that the SMDDL/R {and SMDVL/R} pairs of left/right neurons consistently encodes this control signal, as Fig. \ref{fig:variable_selection}.a shows.

However, as the elimination path is explored further, it is revealed that these well-known neurons can be eliminated from the sparse models and the transition signals can still be reconstructed as shown in \ref{fig:variable_selection}.b.
Indeed, Fig. \ref{fig:variable_selection}.a and and \ref{fig:variable_selection}.b look nearly identical, and Fig. \ref{fig:variable_selection}.c quantifies this using the percentage of false positives and negatives.
Fig. \ref{fig:variable_selection}.c also shows more of the elimination path and when the reconstructions eventually break down.
Fig. \ref{fig:variable_selection}.d and \ref{fig:variable_selection}.e show the how \textbf{K} matrices in equation \ref{eq:encoding} change as neurons are removed.
Taken together, these results reveal previously unidentified neurons that can successfully predict control signals shown to be important to reconstructing the full neural manifold.
However, only rows with names are neurons that have been connected to the stereotyped \textit{C. elegans} connectome and can thus be identified across individuals; rows with numbers cannot be so compared.
In summary, this work identifies sets of neurons previously implicated in transitions and also reveals new candidates for critical actuators of network transitions. 
In addition, the time of encoding is revealed, {which can inform further study}.

\section*{Discussion}

We have presented the first data-driven model that uses a single set of intrinsic dynamics that can reconstruct the multiple behavioral regimes present in a real animal and transitions between them. 
In this study, we have analyzed neuronal recordings from C. elegans that lacked any acutely delivered and time-varying sensory inputs, therefore behavioral transitions are likely internally driven \cite{kato2015global} and governed by stochastic processes \cite{roberts2016stochastic}. 
Here, control signals cannot be inferred simply from sensory neuron activity.
To overcome this challenge, we provide an unsupervised approach for identifying such internally driven control signals and their underlying neuronal identities.
The fact that this controlled linear model accurately reproduces both short and long time-scale dynamics places clear restrictions on the need, specifically the lack thereof, for nonlinearities in this system, and provides hypotheses about the neurons that may contain those nonlinearities as well as their role in the global dynamics of the system. 
In addition, we have embedded this model in a mathematical framework of control, which can be generalized to other organisms or to include hypothesized nonlinearities.

Much excitement has been generated by the availability of the \emph{C. elegans} anatomical connectome, and one aim of data-driven modeling efforts is to produce a functional connectome that can complement the anatomical data.
The DMDc algorithm in this paper is similar to several algorithms in the engineering literature that attempt similar network reconstruction tasks, namely System Identification \cite{ljung2001system}.
One strategy to fully disambiguate the effects of the intrinsic dynamics and the external control signals uses known external perturbations should be applied and the system response measured.
Such perturbations are not generally available in biological systems and thus the data collected are ``uninformative'' \cite{angulo2017fundamental} in the sense that the underlying structure cannot be determined.
Thus, although this is a promising avenue for future work, it is outside the scope of this paper.

A limitation of this model is that it is not generative; it cannot be used to predict a system response that includes transitions to novel stimuli.
To accomplish this, the transition signals must be written as a function of the data.
Step three of our method does this with a linear encoding and demonstrates that the signals can be successfully reconstructed with all neurons to a certain level of accuracy.
If this level of accuracy were sufficient, then the system would be fully linear and an uncontrolled model would produce a good reconstruction, as is clearly not the case.
Recent methods for incorporating nonlinearities into controlled systems (e.g. \cite{brunton2016sparse, williams2015data}) have the potential to create a fully closed-loop feedback system and this is an active area of further research~\cite{morrison2020nonlinear}.

This methodology uses two key hypotheses about this system.
First, although the true biological system includes many thousands of nonlinear interactions, we assume that the leading order or dominant balance dynamics of the dynamical system within certain regimes is simple.
Recent work in the fluids community \cite{batchelor2000introduction} has shown that even when the full global model is perfectly known, almost every region of phase space is well described by a simpler model with fewer or no nonlinear terms.
In the same way, this paper posits that a single set of linear dynamics capture the dominant activity of large regions of phase space.
The second observation is of a time-scale separation between activity within a state, and transitions between states.
This leads to a hypothesis about when complex nonlinearities may be active: sparsely, during transitions between behaviors.
Thus, this methodology is directly applicable when trajectories follow simplified, linear dynamics produced by intrinsic neural dynamics in large regions of phase space, with short periods of complex dynamics.
An area of future work is to explore how these control signals could be produced biologically, and a strength of this method is that the control signals may be well modeled by an intermediate spike-type thresholding of a more complex signal.

A potential criticism of this method is that we have used discrete labeled states in our model, despite ongoing debate regarding how uniform ``states'' in \emph{C. elegans} are across instances, and if they should be subdivided or are simply continuous \cite{szigeti2015searching, liu2018temporal, hums2016regulation}.
We have contributed to this debate by providing evidence that the reversal and turn states in fact appear to be simple and have well-defined initiation signals.
However, the forward ``state'' is much more complex, and breaks the assumptions of our model.
Specifically, the intrinsic dynamics may be different in the forward state as compared to the rest of the phase space, and may be a different linear system as posited in \cite{linderman2019hierarchical} or nonlinear as in \cite{morrison2020nonlinear}.
Related, the ``transition'' into this state may not follow the sparsity assumptions of Algorithm \ref{alg:sparse_residual_analysis}, perhaps due to a continuum of states as opposed to a discrete transition.
Moreover, Kaplan et al. \cite{kaplan2019nested} show that during the forward state many neurons fluctuate at faster timescales, contributing to this complexity.
We argue that this is a strength of this methodology: because this is not a method that can universally approximate arbitrary dynamics, the fact that a state and its transition cannot be reconstructed gives additional information about that state, and about its complexity in relation to other states.
However, it is conceivable that failure to identify control signals during the Forward state, and subsequent low reconstruction quality, is simply due to lack of sufficient data. 
As more neurons are imaged or longer recording times become experimentally feasible, so far undiscovered control signals and neuronal candidates during the Forward state may be revealed.

An alternate approach to modeling complex systems in order to understand structure is to use locally linear models \cite{costa2019adaptive, linderman2016recurrent, linderman2014discovering, linderman2019hierarchical}.
In this methodology, the initial network as described by the matrix $\bf A_k$ is replaced by a new matrix, $\bf A_{k+1}$, at certain change points.
These have achieved great success in reconstructing nonlinear datasets, and in fact can reconstruct arbitrary dynamics given enough change points, and is an active field in machine learning research.
However, it is difficult to interpret what such a replacement of the underlying dynamics would mean biologically, particularly if many separate matrices $\bf A_k$ are required.
On the other hand, the language of control theory from engineering meshes directly with the biological intuition that certain states are initiated by relatively unique signals produced by a small number of neurons.
We propose that our framework for constructing a single, global model of the dynamics of this neural system is promising not only in its ready generalizability to include nonlinearities, but also in its biological interpretability.

We have produced the first, to our knowledge, global data-driven model of both the intrinsic and control dynamics of \emph{C. elegans}.

Here, we provide a framework for the identification of neurons critical in actuating network transitions, which can be tested in future experiments.

\subsection*{References}

\bibliography{Zimmer_paper_arxiv}

\begin{thebibliography}{51}%
\makeatletter
\providecommand \@ifxundefined [1]{%
 \@ifx{#1\undefined}
}%
\providecommand \@ifnum [1]{%
 \ifnum #1\expandafter \@firstoftwo
 \else \expandafter \@secondoftwo
 \fi
}%
\providecommand \@ifx [1]{%
 \ifx #1\expandafter \@firstoftwo
 \else \expandafter \@secondoftwo
 \fi
}%
\providecommand \natexlab [1]{#1}%
\providecommand \enquote  [1]{``#1''}%
\providecommand \bibnamefont  [1]{#1}%
\providecommand \bibfnamefont [1]{#1}%
\providecommand \citenamefont [1]{#1}%
\providecommand \href@noop [0]{\@secondoftwo}%
\providecommand \href [0]{\begingroup \@sanitize@url \@href}%
\providecommand \@href[1]{\@@startlink{#1}\@@href}%
\providecommand \@@href[1]{\endgroup#1\@@endlink}%
\providecommand \@sanitize@url [0]{\catcode `\\12\catcode `\$12\catcode
  `\&12\catcode `\#12\catcode `\^12\catcode `\_12\catcode `\%12\relax}%
\providecommand \@@startlink[1]{}%
\providecommand \@@endlink[0]{}%
\providecommand \url  [0]{\begingroup\@sanitize@url \@url }%
\providecommand \@url [1]{\endgroup\@href {#1}{\urlprefix }}%
\providecommand \urlprefix  [0]{URL }%
\providecommand \Eprint [0]{\href }%
\providecommand \doibase [0]{http://dx.doi.org/}%
\providecommand \selectlanguage [0]{\@gobble}%
\providecommand \bibinfo  [0]{\@secondoftwo}%
\providecommand \bibfield  [0]{\@secondoftwo}%
\providecommand \translation [1]{[#1]}%
\providecommand \BibitemOpen [0]{}%
\providecommand \bibitemStop [0]{}%
\providecommand \bibitemNoStop [0]{.\EOS\space}%
\providecommand \EOS [0]{\spacefactor3000\relax}%
\providecommand \BibitemShut  [1]{\csname bibitem#1\endcsname}%
\let\auto@bib@innerbib\@empty
\bibitem [{\citenamefont {White}\ \emph {et~al.}(1986)\citenamefont {White},
  \citenamefont {Southgate}, \citenamefont {Thomson},\ and\ \citenamefont
  {Brenner}}]{White_paper}%
  \BibitemOpen
  \bibfield  {author} {\bibinfo {author} {\bibfnamefont {J.~G.}\ \bibnamefont
  {White}}, \bibinfo {author} {\bibfnamefont {E.}~\bibnamefont {Southgate}},
  \bibinfo {author} {\bibfnamefont {J.~N.}\ \bibnamefont {Thomson}}, \ and\
  \bibinfo {author} {\bibfnamefont {S.}~\bibnamefont {Brenner}},\ }\href@noop
  {} {\bibfield  {journal} {\bibinfo  {journal} {Philos Trans R Soc Lond B Biol
  Sci}\ }\textbf {\bibinfo {volume} {314}},\ \bibinfo {pages} {1} (\bibinfo
  {year} {1986})}\BibitemShut {NoStop}%
\bibitem [{\citenamefont {Cook}\ \emph {et~al.}(2019)\citenamefont {Cook},
  \citenamefont {Jarrell}, \citenamefont {Brittin}, \citenamefont {Wang},
  \citenamefont {Bloniarz}, \citenamefont {Yakovlev}, \citenamefont {Nguyen},
  \citenamefont {Tang}, \citenamefont {Bayer}, \citenamefont {Duerr} \emph
  {et~al.}}]{cook2019whole}%
  \BibitemOpen
  \bibfield  {author} {\bibinfo {author} {\bibfnamefont {S.~J.}\ \bibnamefont
  {Cook}}, \bibinfo {author} {\bibfnamefont {T.~A.}\ \bibnamefont {Jarrell}},
  \bibinfo {author} {\bibfnamefont {C.~A.}\ \bibnamefont {Brittin}}, \bibinfo
  {author} {\bibfnamefont {Y.}~\bibnamefont {Wang}}, \bibinfo {author}
  {\bibfnamefont {A.~E.}\ \bibnamefont {Bloniarz}}, \bibinfo {author}
  {\bibfnamefont {M.~A.}\ \bibnamefont {Yakovlev}}, \bibinfo {author}
  {\bibfnamefont {K.~C.}\ \bibnamefont {Nguyen}}, \bibinfo {author}
  {\bibfnamefont {L.~T.-H.}\ \bibnamefont {Tang}}, \bibinfo {author}
  {\bibfnamefont {E.~A.}\ \bibnamefont {Bayer}}, \bibinfo {author}
  {\bibfnamefont {J.~S.}\ \bibnamefont {Duerr}},  \emph {et~al.},\ }\href@noop
  {} {\bibfield  {journal} {\bibinfo  {journal} {Nature}\ }\textbf {\bibinfo
  {volume} {571}},\ \bibinfo {pages} {63} (\bibinfo {year} {2019})}\BibitemShut
  {NoStop}%
\bibitem [{\citenamefont {Jarrell}\ \emph {et~al.}(2012)\citenamefont
  {Jarrell}, \citenamefont {Wang}, \citenamefont {Bloniarz}, \citenamefont
  {Brittin}, \citenamefont {Xu}, \citenamefont {Thomson}, \citenamefont
  {Albertson}, \citenamefont {Hall},\ and\ \citenamefont
  {Emmons}}]{jarrell2012connectome}%
  \BibitemOpen
  \bibfield  {author} {\bibinfo {author} {\bibfnamefont {T.~A.}\ \bibnamefont
  {Jarrell}}, \bibinfo {author} {\bibfnamefont {Y.}~\bibnamefont {Wang}},
  \bibinfo {author} {\bibfnamefont {A.~E.}\ \bibnamefont {Bloniarz}}, \bibinfo
  {author} {\bibfnamefont {C.~A.}\ \bibnamefont {Brittin}}, \bibinfo {author}
  {\bibfnamefont {M.}~\bibnamefont {Xu}}, \bibinfo {author} {\bibfnamefont
  {J.~N.}\ \bibnamefont {Thomson}}, \bibinfo {author} {\bibfnamefont {D.~G.}\
  \bibnamefont {Albertson}}, \bibinfo {author} {\bibfnamefont {D.~H.}\
  \bibnamefont {Hall}}, \ and\ \bibinfo {author} {\bibfnamefont {S.~W.}\
  \bibnamefont {Emmons}},\ }\href@noop {} {\bibfield  {journal} {\bibinfo
  {journal} {Science}\ }\textbf {\bibinfo {volume} {337}},\ \bibinfo {pages}
  {437} (\bibinfo {year} {2012})}\BibitemShut {NoStop}%
\bibitem [{\citenamefont {Kato}\ \emph {et~al.}(2015)\citenamefont {Kato},
  \citenamefont {Kaplan}, \citenamefont {Schr{\"o}del}, \citenamefont {Skora},
  \citenamefont {Lindsay}, \citenamefont {Yemini}, \citenamefont {Lockery},\
  and\ \citenamefont {Zimmer}}]{kato2015global}%
  \BibitemOpen
  \bibfield  {author} {\bibinfo {author} {\bibfnamefont {S.}~\bibnamefont
  {Kato}}, \bibinfo {author} {\bibfnamefont {H.~S.}\ \bibnamefont {Kaplan}},
  \bibinfo {author} {\bibfnamefont {T.}~\bibnamefont {Schr{\"o}del}}, \bibinfo
  {author} {\bibfnamefont {S.}~\bibnamefont {Skora}}, \bibinfo {author}
  {\bibfnamefont {T.~H.}\ \bibnamefont {Lindsay}}, \bibinfo {author}
  {\bibfnamefont {E.}~\bibnamefont {Yemini}}, \bibinfo {author} {\bibfnamefont
  {S.}~\bibnamefont {Lockery}}, \ and\ \bibinfo {author} {\bibfnamefont
  {M.}~\bibnamefont {Zimmer}},\ }\href@noop {} {\bibfield  {journal} {\bibinfo
  {journal} {Cell}\ }\textbf {\bibinfo {volume} {163}},\ \bibinfo {pages} {656}
  (\bibinfo {year} {2015})}\BibitemShut {NoStop}%
\bibitem [{\citenamefont {Roberts}\ \emph {et~al.}(2016)\citenamefont
  {Roberts}, \citenamefont {Augustine}, \citenamefont {Lawton}, \citenamefont
  {Lindsay}, \citenamefont {Thiele}, \citenamefont {Izquierdo}, \citenamefont
  {Faumont}, \citenamefont {Lindsay}, \citenamefont {Britton}, \citenamefont
  {Pokala} \emph {et~al.}}]{roberts2016stochastic}%
  \BibitemOpen
  \bibfield  {author} {\bibinfo {author} {\bibfnamefont {W.~M.}\ \bibnamefont
  {Roberts}}, \bibinfo {author} {\bibfnamefont {S.~B.}\ \bibnamefont
  {Augustine}}, \bibinfo {author} {\bibfnamefont {K.~J.}\ \bibnamefont
  {Lawton}}, \bibinfo {author} {\bibfnamefont {T.~H.}\ \bibnamefont {Lindsay}},
  \bibinfo {author} {\bibfnamefont {T.~R.}\ \bibnamefont {Thiele}}, \bibinfo
  {author} {\bibfnamefont {E.~J.}\ \bibnamefont {Izquierdo}}, \bibinfo {author}
  {\bibfnamefont {S.}~\bibnamefont {Faumont}}, \bibinfo {author} {\bibfnamefont
  {R.~A.}\ \bibnamefont {Lindsay}}, \bibinfo {author} {\bibfnamefont {M.~C.}\
  \bibnamefont {Britton}}, \bibinfo {author} {\bibfnamefont {N.}~\bibnamefont
  {Pokala}},  \emph {et~al.},\ }\href@noop {} {\bibfield  {journal} {\bibinfo
  {journal} {Elife}\ }\textbf {\bibinfo {volume} {5}} (\bibinfo {year}
  {2016})}\BibitemShut {NoStop}%
\bibitem [{\citenamefont {Liu}\ \emph {et~al.}(2017)\citenamefont {Liu},
  \citenamefont {Kim},\ and\ \citenamefont {Shlizerman}}]{liu2017functional}%
  \BibitemOpen
  \bibfield  {author} {\bibinfo {author} {\bibfnamefont {H.}~\bibnamefont
  {Liu}}, \bibinfo {author} {\bibfnamefont {J.}~\bibnamefont {Kim}}, \ and\
  \bibinfo {author} {\bibfnamefont {E.}~\bibnamefont {Shlizerman}},\
  }\href@noop {} {\bibfield  {journal} {\bibinfo  {journal} {arXiv preprint
  arXiv:1711.00193}\ } (\bibinfo {year} {2017})}\BibitemShut {NoStop}%
\bibitem [{\citenamefont {Kutz}\ \emph {et~al.}(2016)\citenamefont {Kutz},
  \citenamefont {Brunton}, \citenamefont {Brunton},\ and\ \citenamefont
  {Proctor}}]{kutz2016dynamic}%
  \BibitemOpen
  \bibfield  {author} {\bibinfo {author} {\bibfnamefont {J.~N.}\ \bibnamefont
  {Kutz}}, \bibinfo {author} {\bibfnamefont {S.~L.}\ \bibnamefont {Brunton}},
  \bibinfo {author} {\bibfnamefont {B.~W.}\ \bibnamefont {Brunton}}, \ and\
  \bibinfo {author} {\bibfnamefont {J.~L.}\ \bibnamefont {Proctor}},\
  }\href@noop {} {\emph {\bibinfo {title} {Dynamic mode decomposition:
  data-driven modeling of complex systems}}},\ Vol.\ \bibinfo {volume} {149}\
  (\bibinfo  {publisher} {SIAM},\ \bibinfo {year} {2016})\BibitemShut {NoStop}%
\bibitem [{\citenamefont {Kunert-Graf}\ \emph {et~al.}(2017)\citenamefont
  {Kunert-Graf}, \citenamefont {Shlizerman}, \citenamefont {Walker},\ and\
  \citenamefont {Kutz}}]{kunert2017multistability}%
  \BibitemOpen
  \bibfield  {author} {\bibinfo {author} {\bibfnamefont {J.~M.}\ \bibnamefont
  {Kunert-Graf}}, \bibinfo {author} {\bibfnamefont {E.}~\bibnamefont
  {Shlizerman}}, \bibinfo {author} {\bibfnamefont {A.}~\bibnamefont {Walker}},
  \ and\ \bibinfo {author} {\bibfnamefont {J.~N.}\ \bibnamefont {Kutz}},\
  }\href@noop {} {\bibfield  {journal} {\bibinfo  {journal} {Frontiers in
  computational neuroscience}\ }\textbf {\bibinfo {volume} {11}},\ \bibinfo
  {pages} {53} (\bibinfo {year} {2017})}\BibitemShut {NoStop}%
\bibitem [{\citenamefont {Fieseler}\ \emph {et~al.}(2018)\citenamefont
  {Fieseler}, \citenamefont {Kunert-Graf},\ and\ \citenamefont
  {Kutz}}]{Fieseler_paper1}%
  \BibitemOpen
  \bibfield  {author} {\bibinfo {author} {\bibfnamefont {C.}~\bibnamefont
  {Fieseler}}, \bibinfo {author} {\bibfnamefont {J.}~\bibnamefont
  {Kunert-Graf}}, \ and\ \bibinfo {author} {\bibfnamefont {J.~N.}\ \bibnamefont
  {Kutz}},\ }\href@noop {} {\bibfield  {journal} {\bibinfo  {journal} {Journal
  of Biomechanics}\ }\textbf {\bibinfo {volume} {74}},\ \bibinfo {pages} {1}
  (\bibinfo {year} {2018})}\BibitemShut {NoStop}%
\bibitem [{\citenamefont {Kawano}\ \emph {et~al.}(2011)\citenamefont {Kawano},
  \citenamefont {Po}, \citenamefont {Gao}, \citenamefont {Leung}, \citenamefont
  {Ryu},\ and\ \citenamefont {Zhen}}]{kawano2011imbalancing}%
  \BibitemOpen
  \bibfield  {author} {\bibinfo {author} {\bibfnamefont {T.}~\bibnamefont
  {Kawano}}, \bibinfo {author} {\bibfnamefont {M.~D.}\ \bibnamefont {Po}},
  \bibinfo {author} {\bibfnamefont {S.}~\bibnamefont {Gao}}, \bibinfo {author}
  {\bibfnamefont {G.}~\bibnamefont {Leung}}, \bibinfo {author} {\bibfnamefont
  {W.~S.}\ \bibnamefont {Ryu}}, \ and\ \bibinfo {author} {\bibfnamefont
  {M.}~\bibnamefont {Zhen}},\ }\href@noop {} {\bibfield  {journal} {\bibinfo
  {journal} {Neuron}\ }\textbf {\bibinfo {volume} {72}},\ \bibinfo {pages}
  {572} (\bibinfo {year} {2011})}\BibitemShut {NoStop}%
\bibitem [{\citenamefont {Stephens}\ \emph {et~al.}(2008)\citenamefont
  {Stephens}, \citenamefont {Johnson-Kerner}, \citenamefont {Bialek},\ and\
  \citenamefont {Ryu}}]{stephens2008dimensionality}%
  \BibitemOpen
  \bibfield  {author} {\bibinfo {author} {\bibfnamefont {G.~J.}\ \bibnamefont
  {Stephens}}, \bibinfo {author} {\bibfnamefont {B.}~\bibnamefont
  {Johnson-Kerner}}, \bibinfo {author} {\bibfnamefont {W.}~\bibnamefont
  {Bialek}}, \ and\ \bibinfo {author} {\bibfnamefont {W.~S.}\ \bibnamefont
  {Ryu}},\ }\href@noop {} {\bibfield  {journal} {\bibinfo  {journal} {PLoS
  computational biology}\ }\textbf {\bibinfo {volume} {4}},\ \bibinfo {pages}
  {e1000028} (\bibinfo {year} {2008})}\BibitemShut {NoStop}%
\bibitem [{\citenamefont {Stephens}\ \emph {et~al.}(2011)\citenamefont
  {Stephens}, \citenamefont {de~Mesquita}, \citenamefont {Ryu},\ and\
  \citenamefont {Bialek}}]{stephens2011emergence}%
  \BibitemOpen
  \bibfield  {author} {\bibinfo {author} {\bibfnamefont {G.~J.}\ \bibnamefont
  {Stephens}}, \bibinfo {author} {\bibfnamefont {M.~B.}\ \bibnamefont
  {de~Mesquita}}, \bibinfo {author} {\bibfnamefont {W.~S.}\ \bibnamefont
  {Ryu}}, \ and\ \bibinfo {author} {\bibfnamefont {W.}~\bibnamefont {Bialek}},\
  }\href@noop {} {\bibfield  {journal} {\bibinfo  {journal} {Proceedings of the
  National Academy of Sciences}\ }\textbf {\bibinfo {volume} {108}},\ \bibinfo
  {pages} {7286} (\bibinfo {year} {2011})}\BibitemShut {NoStop}%
\bibitem [{\citenamefont {Nichols}\ \emph {et~al.}(2017)\citenamefont
  {Nichols}, \citenamefont {Eichler}, \citenamefont {Latham},\ and\
  \citenamefont {Zimmer}}]{nichols2017global}%
  \BibitemOpen
  \bibfield  {author} {\bibinfo {author} {\bibfnamefont {A.~L.}\ \bibnamefont
  {Nichols}}, \bibinfo {author} {\bibfnamefont {T.}~\bibnamefont {Eichler}},
  \bibinfo {author} {\bibfnamefont {R.}~\bibnamefont {Latham}}, \ and\ \bibinfo
  {author} {\bibfnamefont {M.}~\bibnamefont {Zimmer}},\ }\href@noop {}
  {\bibfield  {journal} {\bibinfo  {journal} {Science}\ }\textbf {\bibinfo
  {volume} {356}},\ \bibinfo {pages} {eaam6851} (\bibinfo {year}
  {2017})}\BibitemShut {NoStop}%
\bibitem [{\citenamefont {Brennan}\ and\ \citenamefont
  {Proekt}(2019)}]{brennan2019quantitative}%
  \BibitemOpen
  \bibfield  {author} {\bibinfo {author} {\bibfnamefont {C.}~\bibnamefont
  {Brennan}}\ and\ \bibinfo {author} {\bibfnamefont {A.}~\bibnamefont
  {Proekt}},\ }\href@noop {} {\bibfield  {journal} {\bibinfo  {journal}
  {eLife}\ }\textbf {\bibinfo {volume} {8}},\ \bibinfo {pages} {e46814}
  (\bibinfo {year} {2019})}\BibitemShut {NoStop}%
\bibitem [{\citenamefont {Linderman}\ \emph {et~al.}(2019)\citenamefont
  {Linderman}, \citenamefont {Nichols}, \citenamefont {Blei}, \citenamefont
  {Zimmer},\ and\ \citenamefont {Paninski}}]{linderman2019hierarchical}%
  \BibitemOpen
  \bibfield  {author} {\bibinfo {author} {\bibfnamefont {S.~W.}\ \bibnamefont
  {Linderman}}, \bibinfo {author} {\bibfnamefont {A.~L.}\ \bibnamefont
  {Nichols}}, \bibinfo {author} {\bibfnamefont {D.~M.}\ \bibnamefont {Blei}},
  \bibinfo {author} {\bibfnamefont {M.}~\bibnamefont {Zimmer}}, \ and\ \bibinfo
  {author} {\bibfnamefont {L.}~\bibnamefont {Paninski}},\ }\href@noop {}
  {\bibfield  {journal} {\bibinfo  {journal} {bioRxiv}\ ,\ \bibinfo {pages}
  {621540}} (\bibinfo {year} {2019})}\BibitemShut {NoStop}%
\bibitem [{\citenamefont {Zimmer}\ \emph {et~al.}(2009)\citenamefont {Zimmer},
  \citenamefont {Gray}, \citenamefont {Pokala}, \citenamefont {Chang},
  \citenamefont {Karow}, \citenamefont {Marletta}, \citenamefont {Hudson},
  \citenamefont {Morton}, \citenamefont {Chronis},\ and\ \citenamefont
  {Bargmann}}]{zimmer2009neurons}%
  \BibitemOpen
  \bibfield  {author} {\bibinfo {author} {\bibfnamefont {M.}~\bibnamefont
  {Zimmer}}, \bibinfo {author} {\bibfnamefont {J.~M.}\ \bibnamefont {Gray}},
  \bibinfo {author} {\bibfnamefont {N.}~\bibnamefont {Pokala}}, \bibinfo
  {author} {\bibfnamefont {A.~J.}\ \bibnamefont {Chang}}, \bibinfo {author}
  {\bibfnamefont {D.~S.}\ \bibnamefont {Karow}}, \bibinfo {author}
  {\bibfnamefont {M.~A.}\ \bibnamefont {Marletta}}, \bibinfo {author}
  {\bibfnamefont {M.~L.}\ \bibnamefont {Hudson}}, \bibinfo {author}
  {\bibfnamefont {D.~B.}\ \bibnamefont {Morton}}, \bibinfo {author}
  {\bibfnamefont {N.}~\bibnamefont {Chronis}}, \ and\ \bibinfo {author}
  {\bibfnamefont {C.~I.}\ \bibnamefont {Bargmann}},\ }\href@noop {} {\bibfield
  {journal} {\bibinfo  {journal} {Neuron}\ }\textbf {\bibinfo {volume} {61}},\
  \bibinfo {pages} {865} (\bibinfo {year} {2009})}\BibitemShut {NoStop}%
\bibitem [{\citenamefont {Chalasani}\ \emph {et~al.}(2007)\citenamefont
  {Chalasani}, \citenamefont {Chronis}, \citenamefont {Tsunozaki},
  \citenamefont {Gray}, \citenamefont {Ramot}, \citenamefont {Goodman},\ and\
  \citenamefont {Bargmann}}]{chalasani2007dissecting}%
  \BibitemOpen
  \bibfield  {author} {\bibinfo {author} {\bibfnamefont {S.~H.}\ \bibnamefont
  {Chalasani}}, \bibinfo {author} {\bibfnamefont {N.}~\bibnamefont {Chronis}},
  \bibinfo {author} {\bibfnamefont {M.}~\bibnamefont {Tsunozaki}}, \bibinfo
  {author} {\bibfnamefont {J.~M.}\ \bibnamefont {Gray}}, \bibinfo {author}
  {\bibfnamefont {D.}~\bibnamefont {Ramot}}, \bibinfo {author} {\bibfnamefont
  {M.~B.}\ \bibnamefont {Goodman}}, \ and\ \bibinfo {author} {\bibfnamefont
  {C.~I.}\ \bibnamefont {Bargmann}},\ }\href@noop {} {\bibfield  {journal}
  {\bibinfo  {journal} {Nature}\ }\textbf {\bibinfo {volume} {450}},\ \bibinfo
  {pages} {63} (\bibinfo {year} {2007})}\BibitemShut {NoStop}%
\bibitem [{\citenamefont {Chiba}\ and\ \citenamefont
  {Rankin}(1990)}]{chiba1990developmental}%
  \BibitemOpen
  \bibfield  {author} {\bibinfo {author} {\bibfnamefont {C.~M.}\ \bibnamefont
  {Chiba}}\ and\ \bibinfo {author} {\bibfnamefont {C.~H.}\ \bibnamefont
  {Rankin}},\ }\href@noop {} {\bibfield  {journal} {\bibinfo  {journal}
  {Journal of neurobiology}\ }\textbf {\bibinfo {volume} {21}},\ \bibinfo
  {pages} {543} (\bibinfo {year} {1990})}\BibitemShut {NoStop}%
\bibitem [{\citenamefont {Kocabas}\ \emph {et~al.}(2012)\citenamefont
  {Kocabas}, \citenamefont {Shen}, \citenamefont {Guo},\ and\ \citenamefont
  {Ramanathan}}]{kocabas2012controlling}%
  \BibitemOpen
  \bibfield  {author} {\bibinfo {author} {\bibfnamefont {A.}~\bibnamefont
  {Kocabas}}, \bibinfo {author} {\bibfnamefont {C.-H.}\ \bibnamefont {Shen}},
  \bibinfo {author} {\bibfnamefont {Z.~V.}\ \bibnamefont {Guo}}, \ and\
  \bibinfo {author} {\bibfnamefont {S.}~\bibnamefont {Ramanathan}},\
  }\href@noop {} {\bibfield  {journal} {\bibinfo  {journal} {Nature}\ }\textbf
  {\bibinfo {volume} {490}},\ \bibinfo {pages} {273} (\bibinfo {year}
  {2012})}\BibitemShut {NoStop}%
\bibitem [{\citenamefont {Leifer}\ \emph {et~al.}(2011)\citenamefont {Leifer},
  \citenamefont {Fang-Yen}, \citenamefont {Gershow}, \citenamefont {Alkema},\
  and\ \citenamefont {Samuel}}]{leifer2011optogenetic}%
  \BibitemOpen
  \bibfield  {author} {\bibinfo {author} {\bibfnamefont {A.~M.}\ \bibnamefont
  {Leifer}}, \bibinfo {author} {\bibfnamefont {C.}~\bibnamefont {Fang-Yen}},
  \bibinfo {author} {\bibfnamefont {M.}~\bibnamefont {Gershow}}, \bibinfo
  {author} {\bibfnamefont {M.~J.}\ \bibnamefont {Alkema}}, \ and\ \bibinfo
  {author} {\bibfnamefont {A.~D.}\ \bibnamefont {Samuel}},\ }\href@noop {}
  {\bibfield  {journal} {\bibinfo  {journal} {Nature methods}\ }\textbf
  {\bibinfo {volume} {8}},\ \bibinfo {pages} {147} (\bibinfo {year}
  {2011})}\BibitemShut {NoStop}%
\bibitem [{\citenamefont {Nagel}\ \emph {et~al.}(2005)\citenamefont {Nagel},
  \citenamefont {Brauner}, \citenamefont {Liewald}, \citenamefont {Adeishvili},
  \citenamefont {Bamberg},\ and\ \citenamefont {Gottschalk}}]{nagel2005light}%
  \BibitemOpen
  \bibfield  {author} {\bibinfo {author} {\bibfnamefont {G.}~\bibnamefont
  {Nagel}}, \bibinfo {author} {\bibfnamefont {M.}~\bibnamefont {Brauner}},
  \bibinfo {author} {\bibfnamefont {J.~F.}\ \bibnamefont {Liewald}}, \bibinfo
  {author} {\bibfnamefont {N.}~\bibnamefont {Adeishvili}}, \bibinfo {author}
  {\bibfnamefont {E.}~\bibnamefont {Bamberg}}, \ and\ \bibinfo {author}
  {\bibfnamefont {A.}~\bibnamefont {Gottschalk}},\ }\href@noop {} {\bibfield
  {journal} {\bibinfo  {journal} {Current Biology}\ }\textbf {\bibinfo {volume}
  {15}},\ \bibinfo {pages} {2279} (\bibinfo {year} {2005})}\BibitemShut
  {NoStop}%
\bibitem [{\citenamefont {Linderman}\ and\ \citenamefont
  {Adams}(2014)}]{linderman2014discovering}%
  \BibitemOpen
  \bibfield  {author} {\bibinfo {author} {\bibfnamefont {S.}~\bibnamefont
  {Linderman}}\ and\ \bibinfo {author} {\bibfnamefont {R.}~\bibnamefont
  {Adams}},\ }in\ \href@noop {} {\emph {\bibinfo {booktitle} {International
  Conference on Machine Learning}}}\ (\bibinfo {year} {2014})\ pp.\ \bibinfo
  {pages} {1413--1421}\BibitemShut {NoStop}%
\bibitem [{\citenamefont {Linderman}\ \emph {et~al.}(2016)\citenamefont
  {Linderman}, \citenamefont {Miller}, \citenamefont {Adams}, \citenamefont
  {Blei}, \citenamefont {Paninski},\ and\ \citenamefont
  {Johnson}}]{linderman2016recurrent}%
  \BibitemOpen
  \bibfield  {author} {\bibinfo {author} {\bibfnamefont {S.~W.}\ \bibnamefont
  {Linderman}}, \bibinfo {author} {\bibfnamefont {A.~C.}\ \bibnamefont
  {Miller}}, \bibinfo {author} {\bibfnamefont {R.~P.}\ \bibnamefont {Adams}},
  \bibinfo {author} {\bibfnamefont {D.~M.}\ \bibnamefont {Blei}}, \bibinfo
  {author} {\bibfnamefont {L.}~\bibnamefont {Paninski}}, \ and\ \bibinfo
  {author} {\bibfnamefont {M.~J.}\ \bibnamefont {Johnson}},\ }\href@noop {}
  {\bibfield  {journal} {\bibinfo  {journal} {arXiv preprint arXiv:1610.08466}\
  } (\bibinfo {year} {2016})}\BibitemShut {NoStop}%
\bibitem [{\citenamefont {Costa}\ \emph {et~al.}(2019)\citenamefont {Costa},
  \citenamefont {Ahamed},\ and\ \citenamefont {Stephens}}]{costa2019adaptive}%
  \BibitemOpen
  \bibfield  {author} {\bibinfo {author} {\bibfnamefont {A.~C.}\ \bibnamefont
  {Costa}}, \bibinfo {author} {\bibfnamefont {T.}~\bibnamefont {Ahamed}}, \
  and\ \bibinfo {author} {\bibfnamefont {G.~J.}\ \bibnamefont {Stephens}},\
  }\href@noop {} {\bibfield  {journal} {\bibinfo  {journal} {Proceedings of the
  National Academy of Sciences}\ }\textbf {\bibinfo {volume} {116}},\ \bibinfo
  {pages} {1501} (\bibinfo {year} {2019})}\BibitemShut {NoStop}%
\bibitem [{\citenamefont {Kunert}\ \emph {et~al.}(2014)\citenamefont {Kunert},
  \citenamefont {Shlizerman},\ and\ \citenamefont {Kutz}}]{kunert2014low}%
  \BibitemOpen
  \bibfield  {author} {\bibinfo {author} {\bibfnamefont {J.}~\bibnamefont
  {Kunert}}, \bibinfo {author} {\bibfnamefont {E.}~\bibnamefont {Shlizerman}},
  \ and\ \bibinfo {author} {\bibfnamefont {J.~N.}\ \bibnamefont {Kutz}},\
  }\href@noop {} {\bibfield  {journal} {\bibinfo  {journal} {Physical Review
  E}\ }\textbf {\bibinfo {volume} {89}},\ \bibinfo {pages} {052805} (\bibinfo
  {year} {2014})}\BibitemShut {NoStop}%
\bibitem [{\citenamefont {Mujika}\ \emph {et~al.}(2017)\citenamefont {Mujika},
  \citenamefont {Le{\v{s}}kovsk{\`y}}, \citenamefont {{\'A}lvarez},
  \citenamefont {Otaduy},\ and\ \citenamefont {Epelde}}]{mujika2017modeling}%
  \BibitemOpen
  \bibfield  {author} {\bibinfo {author} {\bibfnamefont {A.}~\bibnamefont
  {Mujika}}, \bibinfo {author} {\bibfnamefont {P.}~\bibnamefont
  {Le{\v{s}}kovsk{\`y}}}, \bibinfo {author} {\bibfnamefont {R.}~\bibnamefont
  {{\'A}lvarez}}, \bibinfo {author} {\bibfnamefont {M.~A.}\ \bibnamefont
  {Otaduy}}, \ and\ \bibinfo {author} {\bibfnamefont {G.}~\bibnamefont
  {Epelde}},\ }\href@noop {} {\bibfield  {journal} {\bibinfo  {journal}
  {Frontiers in neuroinformatics}\ }\textbf {\bibinfo {volume} {11}},\ \bibinfo
  {pages} {71} (\bibinfo {year} {2017})}\BibitemShut {NoStop}%
\bibitem [{\citenamefont {Costalago-Meruelo}\ \emph {et~al.}(2018)\citenamefont
  {Costalago-Meruelo}, \citenamefont {Machado}, \citenamefont {Appiah},
  \citenamefont {Mujika}, \citenamefont {Leskovsky}, \citenamefont {Alvarez},
  \citenamefont {Epelde},\ and\ \citenamefont
  {McGinnity}}]{costalago2018emulation}%
  \BibitemOpen
  \bibfield  {author} {\bibinfo {author} {\bibfnamefont {A.}~\bibnamefont
  {Costalago-Meruelo}}, \bibinfo {author} {\bibfnamefont {P.}~\bibnamefont
  {Machado}}, \bibinfo {author} {\bibfnamefont {K.}~\bibnamefont {Appiah}},
  \bibinfo {author} {\bibfnamefont {A.}~\bibnamefont {Mujika}}, \bibinfo
  {author} {\bibfnamefont {P.}~\bibnamefont {Leskovsky}}, \bibinfo {author}
  {\bibfnamefont {R.}~\bibnamefont {Alvarez}}, \bibinfo {author} {\bibfnamefont
  {G.}~\bibnamefont {Epelde}}, \ and\ \bibinfo {author} {\bibfnamefont {T.~M.}\
  \bibnamefont {McGinnity}},\ }\href@noop {} {\bibfield  {journal} {\bibinfo
  {journal} {Neurocomputing}\ }\textbf {\bibinfo {volume} {290}},\ \bibinfo
  {pages} {60} (\bibinfo {year} {2018})}\BibitemShut {NoStop}%
\bibitem [{\citenamefont {Izquierdo}(2018)}]{izquierdo2018role}%
  \BibitemOpen
  \bibfield  {author} {\bibinfo {author} {\bibfnamefont {E.~J.}\ \bibnamefont
  {Izquierdo}},\ }\href@noop {} {\bibfield  {journal} {\bibinfo  {journal}
  {Current Opinion in Systems Biology}\ } (\bibinfo {year} {2018})}\BibitemShut
  {NoStop}%
\bibitem [{\citenamefont {Gleeson}\ \emph {et~al.}(2018)\citenamefont
  {Gleeson}, \citenamefont {Lung}, \citenamefont {Grosu}, \citenamefont
  {Hasani},\ and\ \citenamefont {Larson}}]{gleeson2018c302}%
  \BibitemOpen
  \bibfield  {author} {\bibinfo {author} {\bibfnamefont {P.}~\bibnamefont
  {Gleeson}}, \bibinfo {author} {\bibfnamefont {D.}~\bibnamefont {Lung}},
  \bibinfo {author} {\bibfnamefont {R.}~\bibnamefont {Grosu}}, \bibinfo
  {author} {\bibfnamefont {R.}~\bibnamefont {Hasani}}, \ and\ \bibinfo {author}
  {\bibfnamefont {S.~D.}\ \bibnamefont {Larson}},\ }\href@noop {} {\bibfield
  {journal} {\bibinfo  {journal} {Philosophical Transactions of the Royal
  Society B: Biological Sciences}\ }\textbf {\bibinfo {volume} {373}},\
  \bibinfo {pages} {20170379} (\bibinfo {year} {2018})}\BibitemShut {NoStop}%
\bibitem [{\citenamefont {Boyle}\ \emph {et~al.}(2012)\citenamefont {Boyle},
  \citenamefont {Berri},\ and\ \citenamefont {Cohen}}]{boyle2012gait}%
  \BibitemOpen
  \bibfield  {author} {\bibinfo {author} {\bibfnamefont {J.~H.}\ \bibnamefont
  {Boyle}}, \bibinfo {author} {\bibfnamefont {S.}~\bibnamefont {Berri}}, \ and\
  \bibinfo {author} {\bibfnamefont {N.}~\bibnamefont {Cohen}},\ }\href@noop {}
  {\bibfield  {journal} {\bibinfo  {journal} {Frontiers in computational
  neuroscience}\ }\textbf {\bibinfo {volume} {6}},\ \bibinfo {pages} {10}
  (\bibinfo {year} {2012})}\BibitemShut {NoStop}%
\bibitem [{\citenamefont {Proctor}\ \emph {et~al.}(2016)\citenamefont
  {Proctor}, \citenamefont {Brunton},\ and\ \citenamefont
  {Kutz}}]{proctor2016dynamic}%
  \BibitemOpen
  \bibfield  {author} {\bibinfo {author} {\bibfnamefont {J.~L.}\ \bibnamefont
  {Proctor}}, \bibinfo {author} {\bibfnamefont {S.~L.}\ \bibnamefont
  {Brunton}}, \ and\ \bibinfo {author} {\bibfnamefont {J.~N.}\ \bibnamefont
  {Kutz}},\ }\href@noop {} {\bibfield  {journal} {\bibinfo  {journal} {SIAM
  Journal on Applied Dynamical Systems}\ }\textbf {\bibinfo {volume} {15}},\
  \bibinfo {pages} {142} (\bibinfo {year} {2016})}\BibitemShut {NoStop}%
\bibitem [{\citenamefont {Fieseler}(2019)}]{this_code_github}%
  \BibitemOpen
  \bibfield  {author} {\bibinfo {author} {\bibfnamefont {C.}~\bibnamefont
  {Fieseler}},\ }\href@noop {} {\enquote {\bibinfo {title} {Celegansmodel},}\ }
  (\bibinfo {year} {2019})\BibitemShut {NoStop}%
\bibitem [{\citenamefont {Jewell}\ and\ \citenamefont
  {Witten}(2018)}]{jewell2018exact}%
  \BibitemOpen
  \bibfield  {author} {\bibinfo {author} {\bibfnamefont {S.}~\bibnamefont
  {Jewell}}\ and\ \bibinfo {author} {\bibfnamefont {D.}~\bibnamefont
  {Witten}},\ }\href@noop {} {\bibfield  {journal} {\bibinfo  {journal} {The
  annals of applied statistics}\ }\textbf {\bibinfo {volume} {12}},\ \bibinfo
  {pages} {2457} (\bibinfo {year} {2018})}\BibitemShut {NoStop}%
\bibitem [{\citenamefont {Donoho}(2006)}]{donoho2006most}%
  \BibitemOpen
  \bibfield  {author} {\bibinfo {author} {\bibfnamefont {D.~L.}\ \bibnamefont
  {Donoho}},\ }\href@noop {} {\bibfield  {journal} {\bibinfo  {journal}
  {Communications on Pure and Applied Mathematics: A Journal Issued by the
  Courant Institute of Mathematical Sciences}\ }\textbf {\bibinfo {volume}
  {59}},\ \bibinfo {pages} {797} (\bibinfo {year} {2006})}\BibitemShut
  {NoStop}%
\bibitem [{\citenamefont {Su}\ \emph {et~al.}(2017)\citenamefont {Su},
  \citenamefont {Bogdan}, \citenamefont {Candes} \emph {et~al.}}]{su2017false}%
  \BibitemOpen
  \bibfield  {author} {\bibinfo {author} {\bibfnamefont {W.}~\bibnamefont
  {Su}}, \bibinfo {author} {\bibfnamefont {M.}~\bibnamefont {Bogdan}}, \bibinfo
  {author} {\bibfnamefont {E.}~\bibnamefont {Candes}},  \emph {et~al.},\
  }\href@noop {} {\bibfield  {journal} {\bibinfo  {journal} {The Annals of
  Statistics}\ }\textbf {\bibinfo {volume} {45}},\ \bibinfo {pages} {2133}
  (\bibinfo {year} {2017})}\BibitemShut {NoStop}%
\bibitem [{\citenamefont {Brunton}\ \emph {et~al.}(2016)\citenamefont
  {Brunton}, \citenamefont {Proctor},\ and\ \citenamefont
  {Kutz}}]{brunton2016sparse}%
  \BibitemOpen
  \bibfield  {author} {\bibinfo {author} {\bibfnamefont {S.~L.}\ \bibnamefont
  {Brunton}}, \bibinfo {author} {\bibfnamefont {J.~L.}\ \bibnamefont
  {Proctor}}, \ and\ \bibinfo {author} {\bibfnamefont {J.~N.}\ \bibnamefont
  {Kutz}},\ }\href@noop {} {\bibfield  {journal} {\bibinfo  {journal}
  {IFAC-PapersOnLine}\ }\textbf {\bibinfo {volume} {49}},\ \bibinfo {pages}
  {710} (\bibinfo {year} {2016})}\BibitemShut {NoStop}%
\bibitem [{\citenamefont {Zhang}\ and\ \citenamefont
  {Schaeffer}(2018)}]{zhang2018convergence}%
  \BibitemOpen
  \bibfield  {author} {\bibinfo {author} {\bibfnamefont {L.}~\bibnamefont
  {Zhang}}\ and\ \bibinfo {author} {\bibfnamefont {H.}~\bibnamefont
  {Schaeffer}},\ }\href@noop {} {\bibfield  {journal} {\bibinfo  {journal}
  {arXiv preprint arXiv:1805.06445}\ } (\bibinfo {year} {2018})}\BibitemShut
  {NoStop}%
\bibitem [{\citenamefont {Zheng}\ \emph {et~al.}(2018)\citenamefont {Zheng},
  \citenamefont {Askham}, \citenamefont {Brunton}, \citenamefont {Kutz},\ and\
  \citenamefont {Aravkin}}]{zheng2018unified}%
  \BibitemOpen
  \bibfield  {author} {\bibinfo {author} {\bibfnamefont {P.}~\bibnamefont
  {Zheng}}, \bibinfo {author} {\bibfnamefont {T.}~\bibnamefont {Askham}},
  \bibinfo {author} {\bibfnamefont {S.~L.}\ \bibnamefont {Brunton}}, \bibinfo
  {author} {\bibfnamefont {J.~N.}\ \bibnamefont {Kutz}}, \ and\ \bibinfo
  {author} {\bibfnamefont {A.~Y.}\ \bibnamefont {Aravkin}},\ }\href@noop {}
  {\bibfield  {journal} {\bibinfo  {journal} {IEEE Access}\ }\textbf {\bibinfo
  {volume} {7}},\ \bibinfo {pages} {1404} (\bibinfo {year} {2018})}\BibitemShut
  {NoStop}%
\bibitem [{\citenamefont {Tibshirani}(1996)}]{tibshirani1996regression}%
  \BibitemOpen
  \bibfield  {author} {\bibinfo {author} {\bibfnamefont {R.}~\bibnamefont
  {Tibshirani}},\ }\href@noop {} {\bibfield  {journal} {\bibinfo  {journal}
  {Journal of the Royal Statistical Society: Series B (Methodological)}\
  }\textbf {\bibinfo {volume} {58}},\ \bibinfo {pages} {267} (\bibinfo {year}
  {1996})}\BibitemShut {NoStop}%
\bibitem [{\citenamefont {Szigeti}\ \emph {et~al.}(2015)\citenamefont
  {Szigeti}, \citenamefont {Deogade},\ and\ \citenamefont
  {Webb}}]{szigeti2015searching}%
  \BibitemOpen
  \bibfield  {author} {\bibinfo {author} {\bibfnamefont {B.}~\bibnamefont
  {Szigeti}}, \bibinfo {author} {\bibfnamefont {A.}~\bibnamefont {Deogade}}, \
  and\ \bibinfo {author} {\bibfnamefont {B.}~\bibnamefont {Webb}},\ }\href@noop
  {} {\bibfield  {journal} {\bibinfo  {journal} {Journal of the Royal Society
  Interface}\ }\textbf {\bibinfo {volume} {12}},\ \bibinfo {pages} {20150899}
  (\bibinfo {year} {2015})}\BibitemShut {NoStop}%
\bibitem [{\citenamefont {Liu}\ \emph {et~al.}(2018)\citenamefont {Liu},
  \citenamefont {Sharma}, \citenamefont {Shaevitz},\ and\ \citenamefont
  {Leifer}}]{liu2018temporal}%
  \BibitemOpen
  \bibfield  {author} {\bibinfo {author} {\bibfnamefont {M.}~\bibnamefont
  {Liu}}, \bibinfo {author} {\bibfnamefont {A.~K.}\ \bibnamefont {Sharma}},
  \bibinfo {author} {\bibfnamefont {J.~W.}\ \bibnamefont {Shaevitz}}, \ and\
  \bibinfo {author} {\bibfnamefont {A.~M.}\ \bibnamefont {Leifer}},\
  }\href@noop {} {\bibfield  {journal} {\bibinfo  {journal} {Elife}\ }\textbf
  {\bibinfo {volume} {7}},\ \bibinfo {pages} {e36419} (\bibinfo {year}
  {2018})}\BibitemShut {NoStop}%
\bibitem [{\citenamefont {Kaplan}\ \emph {et~al.}(2019)\citenamefont {Kaplan},
  \citenamefont {Thula}, \citenamefont {Khoss},\ and\ \citenamefont
  {Zimmer}}]{kaplan2019nested}%
  \BibitemOpen
  \bibfield  {author} {\bibinfo {author} {\bibfnamefont {H.~S.}\ \bibnamefont
  {Kaplan}}, \bibinfo {author} {\bibfnamefont {O.~S.}\ \bibnamefont {Thula}},
  \bibinfo {author} {\bibfnamefont {N.}~\bibnamefont {Khoss}}, \ and\ \bibinfo
  {author} {\bibfnamefont {M.}~\bibnamefont {Zimmer}},\ }\href@noop {}
  {\bibfield  {journal} {\bibinfo  {journal} {Neuron}\ } (\bibinfo {year}
  {2019})}\BibitemShut {NoStop}%
\bibitem [{\citenamefont {Zhen}\ and\ \citenamefont
  {Samuel}(2015)}]{zhen2015c}%
  \BibitemOpen
  \bibfield  {author} {\bibinfo {author} {\bibfnamefont {M.}~\bibnamefont
  {Zhen}}\ and\ \bibinfo {author} {\bibfnamefont {A.~D.}\ \bibnamefont
  {Samuel}},\ }\href@noop {} {\bibfield  {journal} {\bibinfo  {journal}
  {Current opinion in neurobiology}\ }\textbf {\bibinfo {volume} {33}},\
  \bibinfo {pages} {117} (\bibinfo {year} {2015})}\BibitemShut {NoStop}%
\bibitem [{\citenamefont {Iino}\ and\ \citenamefont
  {Yoshida}(2009)}]{iino2009parallel}%
  \BibitemOpen
  \bibfield  {author} {\bibinfo {author} {\bibfnamefont {Y.}~\bibnamefont
  {Iino}}\ and\ \bibinfo {author} {\bibfnamefont {K.}~\bibnamefont {Yoshida}},\
  }\href@noop {} {\bibfield  {journal} {\bibinfo  {journal} {Journal of
  Neuroscience}\ }\textbf {\bibinfo {volume} {29}},\ \bibinfo {pages} {5370}
  (\bibinfo {year} {2009})}\BibitemShut {NoStop}%
\bibitem [{\citenamefont {Luo}\ \emph {et~al.}(2006)\citenamefont {Luo},
  \citenamefont {Clark}, \citenamefont {Biron}, \citenamefont {Mahadevan},\
  and\ \citenamefont {Samuel}}]{luo2006sensorimotor}%
  \BibitemOpen
  \bibfield  {author} {\bibinfo {author} {\bibfnamefont {L.}~\bibnamefont
  {Luo}}, \bibinfo {author} {\bibfnamefont {D.~A.}\ \bibnamefont {Clark}},
  \bibinfo {author} {\bibfnamefont {D.}~\bibnamefont {Biron}}, \bibinfo
  {author} {\bibfnamefont {L.}~\bibnamefont {Mahadevan}}, \ and\ \bibinfo
  {author} {\bibfnamefont {A.~D.}\ \bibnamefont {Samuel}},\ }\href@noop {}
  {\bibfield  {journal} {\bibinfo  {journal} {Journal of experimental biology}\
  }\textbf {\bibinfo {volume} {209}},\ \bibinfo {pages} {4652} (\bibinfo {year}
  {2006})}\BibitemShut {NoStop}%
\bibitem [{\citenamefont {Ljung}(2001)}]{ljung2001system}%
  \BibitemOpen
  \bibfield  {author} {\bibinfo {author} {\bibfnamefont {L.}~\bibnamefont
  {Ljung}},\ }\href@noop {} {\bibfield  {journal} {\bibinfo  {journal} {Wiley
  Encyclopedia of Electrical and Electronics Engineering}\ } (\bibinfo {year}
  {2001})}\BibitemShut {NoStop}%
\bibitem [{\citenamefont {Angulo}\ \emph {et~al.}(2017)\citenamefont {Angulo},
  \citenamefont {Moreno}, \citenamefont {Lippner}, \citenamefont
  {Barab{\'a}si},\ and\ \citenamefont {Liu}}]{angulo2017fundamental}%
  \BibitemOpen
  \bibfield  {author} {\bibinfo {author} {\bibfnamefont {M.~T.}\ \bibnamefont
  {Angulo}}, \bibinfo {author} {\bibfnamefont {J.~A.}\ \bibnamefont {Moreno}},
  \bibinfo {author} {\bibfnamefont {G.}~\bibnamefont {Lippner}}, \bibinfo
  {author} {\bibfnamefont {A.-L.}\ \bibnamefont {Barab{\'a}si}}, \ and\
  \bibinfo {author} {\bibfnamefont {Y.-Y.}\ \bibnamefont {Liu}},\ }\href@noop
  {} {\bibfield  {journal} {\bibinfo  {journal} {Journal of the Royal Society
  Interface}\ }\textbf {\bibinfo {volume} {14}},\ \bibinfo {pages} {20160966}
  (\bibinfo {year} {2017})}\BibitemShut {NoStop}%
\bibitem [{\citenamefont {Williams}\ \emph {et~al.}(2015)\citenamefont
  {Williams}, \citenamefont {Kevrekidis},\ and\ \citenamefont
  {Rowley}}]{williams2015data}%
  \BibitemOpen
  \bibfield  {author} {\bibinfo {author} {\bibfnamefont {M.~O.}\ \bibnamefont
  {Williams}}, \bibinfo {author} {\bibfnamefont {I.~G.}\ \bibnamefont
  {Kevrekidis}}, \ and\ \bibinfo {author} {\bibfnamefont {C.~W.}\ \bibnamefont
  {Rowley}},\ }\href@noop {} {\bibfield  {journal} {\bibinfo  {journal}
  {Journal of Nonlinear Science}\ }\textbf {\bibinfo {volume} {25}},\ \bibinfo
  {pages} {1307} (\bibinfo {year} {2015})}\BibitemShut {NoStop}%
\bibitem [{\citenamefont {Morrison}(lder)}]{megan_placeholder}%
  \BibitemOpen
  \bibfield  {author} {\bibinfo {author} {\bibfnamefont {M.}~\bibnamefont
  {Morrison}},\ }\href@noop {} {\bibfield  {journal} {\bibinfo  {journal}
  {Placeholder}\ } (\bibinfo {year} {Placeholder})}\BibitemShut {NoStop}%
\bibitem [{\citenamefont {Batchelor}\ and\ \citenamefont
  {Batchelor}(2000)}]{batchelor2000introduction}%
  \BibitemOpen
  \bibfield  {author} {\bibinfo {author} {\bibfnamefont {C.~K.}\ \bibnamefont
  {Batchelor}}\ and\ \bibinfo {author} {\bibfnamefont {G.}~\bibnamefont
  {Batchelor}},\ }\href@noop {} {\emph {\bibinfo {title} {An introduction to
  fluid dynamics}}}\ (\bibinfo  {publisher} {Cambridge university press},\
  \bibinfo {year} {2000})\BibitemShut {NoStop}%
\bibitem [{\citenamefont {Hums}\ \emph {et~al.}(2016)\citenamefont {Hums},
  \citenamefont {Riedl}, \citenamefont {Mende}, \citenamefont {Kato},
  \citenamefont {Kaplan}, \citenamefont {Latham}, \citenamefont {Sonntag},
  \citenamefont {Traunm{\"u}ller},\ and\ \citenamefont
  {Zimmer}}]{hums2016regulation}%
  \BibitemOpen
  \bibfield  {author} {\bibinfo {author} {\bibfnamefont {I.}~\bibnamefont
  {Hums}}, \bibinfo {author} {\bibfnamefont {J.}~\bibnamefont {Riedl}},
  \bibinfo {author} {\bibfnamefont {F.}~\bibnamefont {Mende}}, \bibinfo
  {author} {\bibfnamefont {S.}~\bibnamefont {Kato}}, \bibinfo {author}
  {\bibfnamefont {H.~S.}\ \bibnamefont {Kaplan}}, \bibinfo {author}
  {\bibfnamefont {R.}~\bibnamefont {Latham}}, \bibinfo {author} {\bibfnamefont
  {M.}~\bibnamefont {Sonntag}}, \bibinfo {author} {\bibfnamefont
  {L.}~\bibnamefont {Traunm{\"u}ller}}, \ and\ \bibinfo {author} {\bibfnamefont
  {M.}~\bibnamefont {Zimmer}},\ }\href@noop {} {\bibfield  {journal} {\bibinfo
  {journal} {Elife}\ }\textbf {\bibinfo {volume} {5}},\ \bibinfo {pages}
  {e14116} (\bibinfo {year} {2016})}\BibitemShut {NoStop}%
\end{thebibliography}%

%


\end{document}